\documentclass[aps,amsmath,amssymb,twocolumn]{revtex4-1}

\usepackage{epsfig}
\usepackage{graphicx}
\usepackage{color}
\usepackage{mathrsfs}
\usepackage{amsfonts}
\usepackage[english]{babel}
\usepackage{color}

\begin{document}

\title{\large \bf Spin-reorientation critical dynamics in two-dimensional XY model with a domain wall}

\author{X. W. Lei$^{1}$, N. J. Zhou$^{2, *}$, Y. Y. He$^{3}$ and B. Zheng$^{3,4,}$}
\email[Corresponding author:~]{zhounengji@hznu.edu.cn}
\email[\\Corresponding author:~]{zhengbo@hznu.edu.cn}
\affiliation{$^1$Institute of electronic information and automation, Aba Teachers university, Wenchuan 623002, People's Republic of China\\
$^2$Department of Physics, Hangzhou Normal University, Hangzhou 311121, People's Republic of China\\
$^3$Department of Physics, Zhejiang University, Hangzhou 310027, People's Republic of China \\
$^4$Collaborative Innovation Center of Advanced Microstructures, Nanjing University, Nanjing 210093, People's Republic of China
}

\begin{abstract}
In recent years, static and dynamic properties of non-$180^\circ$ domain walls in magnetic materials have attracted a great deal of interest.
In this paper, spin-reorientation critical dynamics in the two-dimensional XY model is investigated with Monte Carlo simulations and theoretical analyses based on the Langevin equation. At the Kosterlitz-Thouless phase transition, dynamic scaling behaviors of the magnetization and the two-time correlation function are carefully analyzed, and critical exponents are accurately determined. When the initial value of the angle between adjacent domains is slightly lower than $\pi$, a critical exponent is introduced to characterize the abnormal power-law increase of the magnetization in the horizontal direction inside the domain interface, which is measured to be $\psi=0.0568(8)$. Besides, the relation $\psi=\eta/2z$ is analytically deduced from the Langevin dynamics in the long-wavelength
approximation, well consistent with numerical results.
\end{abstract}

%\showpacs{64.60.Ht, 68.35.Rh, 05.10.Ln}

\maketitle

\section{Introduction}
In the past years, much effort of physicists has been devoted to the understanding of the domain-wall dynamics of ferroic materials
(ferroelectrics, ferromagnets, ferroelastics) in both experiments and theories \cite{lem98,yan99,gau01,cha00,li04,yam07, kle07,cat12}, due to the possible applications in high-density magnetic memories, spin logic devices, and shift registers by means of switching and detecting the polarization orientations of the domains \cite{all02, all05,par08,hay08}. The dynamic properties of domain walls in the macroscopic, mesoscopic, and microscopic scales have been investigated with different numerical methods, such as the Edwards-Wilkinson equation with quenched disorder \cite{cha00,kol05, due05}, Monte Carlo method in the Ising-type lattice models \cite{now98,sep01,xi05,zho09}, and Landau-Lifshitz-Gilbert equation in the Heisenberg-like models \cite{ohe06,gou10,mor17,xio18}.

Driven by a constant external field in the presence of the quenched disorder, a pinning-depinning dynamic transition occurs at zero temperature, separating the regimes of static pinning and friction-limited viscous sliding \cite{fer13,si16,xio18}.  At low temperatures, the sharp depinning transition is softened, and a thermally activated creep motion appears \cite{zho14,gen16,par17,jin18}. Under an oscillating driving field, the situation of domain-wall motion becomes more complicated. Four dynamic states (relaxation, creep, sliding and switching) and dynamic phase transition between them have been found in ultrathin ferromagnetic and ferroelectric films \cite{bra05,zho10}. Recently, domain-wall motions induced by the spin-polarized currents and spin waves have attracted much attention as well \cite{li04,ohe06,log10,wan12}.

\begin{figure*}[p]
\epsfysize=9cm \epsfclipoff \fboxsep=0pt
\setlength{\unitlength}{1.cm}
\begin{picture}(10,9)(0,0)
\put(0,0){{\epsffile{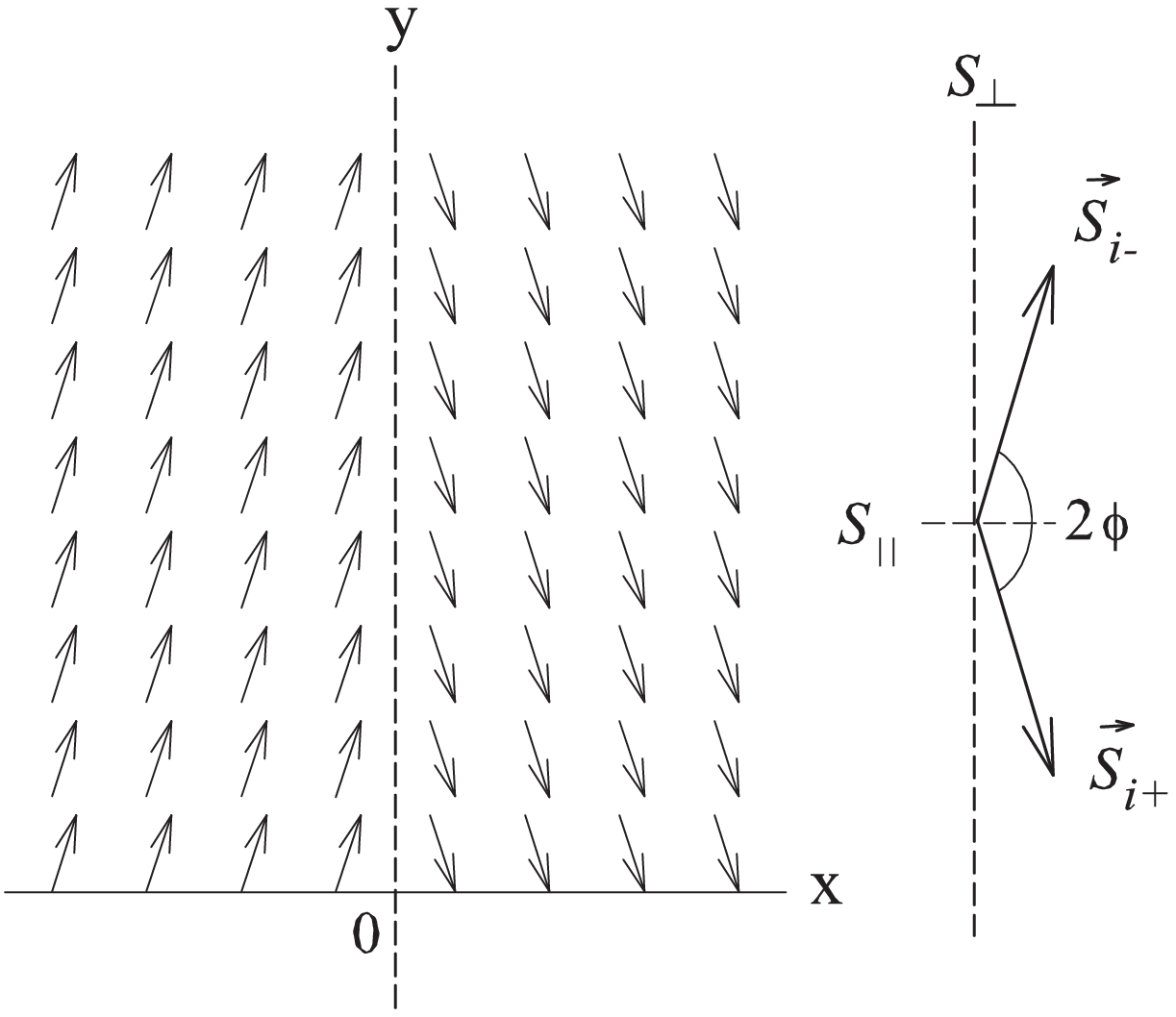}}}
\end{picture}
\caption{The initial spin configuration of a semi-ordered state is shown for the two-dimensional XY model. Spins of both sides form an angle of $2\phi\in[0,\pi]$. The vertical and horizontal components of the spins, shown in the enlargement of two arrows on the right, are denoted by $S_\perp$ and $S_\|$ respectively.}
\label{f1}
\end{figure*}

The structure of the domain wall is very important to the topic of the domain-wall dynamics, which is  of Ne\'el/Bloch type where the magnetization rotates in/out of plane across the domain wall. Thanks to the spin-polarized scanning tunnelling microscopy and spin-polarized low-energy electron microscopy, direct determination of the domain-wall structure is possible by imaging the local in-plane magnetisation direction \cite{mec09,che13,tet15}. To character the domain arrangements, an angle between polarization directions in adjacent domains is introduced as depicted in Fig.~\ref{f1}, which is equal to $\pi$ ($180^\circ$) for the simple case of an easy-axis magnetisation. In recent experiments, spin configurations with non-$180^\circ$ domain walls have also been found, but relevant theoretical explanations are still primitive \cite{nel11, cat12,goo15,tru16}. Very recently, a spin-reorientation transition has been revealed in epitaxial NdCo$_5$ thin film with temperature-dependent domain-wall orientations \cite{sei17}. However, the dynamic properties have not been touched yet.

On the other hand, dynamic relaxation of a single domain wall at a standard ordered-disordered phase transition has already been concerned in magnetic systems \cite{zho08,he09,jin18}. Understanding such a domain-wall dynamics is theoretically and practically important. Theoretically, it is quite interesting to investigate the non-equilibrium critical dynamics starting from the semi-ordered state possessing a single domain wall, in comparison with that starting from the ordered or random state. Practically, predicting and controlling the movements of the domain walls play a crucial role in designing new classes of magnetic devices. Moreover, the dynamic approaches can be developed to study the pinning-depinning and other dynamic phase transitions of domain walls at zero or low temperatures, understand the non-stationary properties of the dynamic systems, and determine the static and dynamic exponents as well as the transition points \cite{zho09,zho14,si16,xio18}. However, most of efforts were focused on the $180^\circ$ domain wall in earlier studies of domain dynamics, and the dynamics of non-$180^\circ$ one was rarely referred.

In this paper, we aim to study the spin-reorientation dynamics with a domain wall at the Kosterlitz-Thouless (KT) phase transition,
taking the two-dimensional ($2$D) XY model as an example. To be specific, the dynamic relaxation starting from the semi-ordered state which consists of two fully-ordered domains with different spin orientations is carefully investigated with Monte Carlo simulations, in comparison with the theoretical analysis based on the Langevin equation in the long-wavelength approximation \cite{kim97, ber01, kap07}. Different values other than $\pi$ are set to the angle $2\phi$ between spins in the two domains, and an abnormal increase can be found in the time evolution of magnetization when the angle is very close to $\pi$, e.g., $0.988\pi$. The critical scaling behavior of the magnetization is worked out analytically, in comparison with the results obtained by Monte Carlo simulations. In Sec. II, the models and scaling analysis are described. And in Secs. III and IV, Monte Carlo simulations and theoretical analysis on the critical behaviors of the magnetization are presented, respectively. Finally, Sec. V includes the conclusions.

\section{Model and scaling analysis}
Two-dimension XY model is one of the simplest models for magnetic
materials, exhibiting a Kosterlitz-Thouless phase transition. The
Hamiltonian is given by
\begin{equation}
-\frac{1}{kT}\mathcal{H}= K  \sum_{<ij>}  \vec{S_i}\cdot \vec{S_j} ,
\label{equ5}
\end{equation}
where $\vec{S_i} =(S_{i,x},S_{i,y})$ is a planar unit vector at the site $i$ in a two-dimensional lattice, the sum is over the nearest neighbors, and $K$ represents
the inverse temperature $1/T$. In this paper, we investigate the relaxation dynamics of the domain walls at the KT phase transition. Since the critical temperature $T_{\rm c}$ is reported to be between $0.89$ and $0.90$ \cite{tom02}, and the system remains critical in the low-temperature phase, we set the temperature $T = 0.89$ in the numerical simulations.  Following Refs.\cite{bra00,he09}, we adopt the``heat-bath'' algorithm with a standard single-spin flip, in which a trial move is accepted with the probability $1/[1+\exp(\delta E/T)]$, where $\delta E$ is the energy change associated with the move.

A rectangular lattice is used in this work with the linear size $2L$ in the $x$ direction and $L$ in the $y$ direction.
Periodic boundary conditions are used in both directions. The semi-ordered state with a perfect domain wall is built by two
ordered sublattices $L^2$, in which all spins of $\vec{S}_{i-}$ in the same orientation on the left side and those of $\vec{S}_{i+}$ on the right side form an angle of $2\phi$ as depicted in Fig.~\ref{f1}. For convenience, we reset the $x$-axis such that the domain wall between the positive and negative spins is located at $x = 0$. So the $x$ coordinate of a lattice site is a half-integer.

Due to the semi-ordered initial state, the time evolution of the dynamic system is inhomogeneous in the $x$ direction. Therefore the magnetization and its second moment should be calculated as functions of $x$ and $t$,
\begin{equation}
\vec{M}^{(k)}(t,x) = \frac{1}{L^k} \left \langle \left [\sum^L_{y =
1}\vec{S}_{xy}(t)\right ]^k \right \rangle,  \quad k = 1,2,
\label{equ10}
\end{equation}
where $\vec{S}_{xy}(t)$ is the spin at time $t$ on site $(x,y)$, $L$ is the lattice size in the $y$ direction, and $\langle\cdots\rangle$ represents the statistical
average. For convenience, we also use $\vec{M}(t,x) \equiv \vec{M}^{(1)}(t,x)$ to denote the magnetization. Same as the spin $\vec{S}_{xy}$, $\vec{M}(t,x)$ also consists of two orthogonal components $M_\perp(t,x)$ and $M_\|(t,x)$ in the vertical and horizontal directions, respectively. For example, one has the initial values $M_\perp(0,x)=\pm 1$ and $M_\|(0,x)=0$ of the $180^\circ$ domain wall.

Besides, the two-time correlation function is introduced to describe the pure
temporal correlation of the domain interface at different times,
\begin{equation}
C(t',t,x) = \frac{1}{L}\left \langle \sum^L_{y
=1}\vec{S}_{xy}(t')\cdot\vec{S}_{xy}(t)\right \rangle -\vec{M}(t',x)\cdot\vec{M}(t,x),
\label{equ20}
\end{equation}
where $t'$ and $t$ denote the waiting and observation time, respectively, and one has $t>t'$ in general.

In the critical regime $T\leq T_{\rm c}$, general scaling arguments lead to the scaling
form of the dynamics far away from the equilibrium state. Similar with that in the Ising model,
the vertical component $M_\perp$ exhibits as
\begin{equation}
M_\perp^{(k)}(t,x,L)=\xi(t)^{-k\eta /2} \widetilde{M}_\perp^{(k)}(\xi(t)/x, \xi(t)/L),
\label{equ25}
\end{equation}
where $\eta$ is the static exponent, $k = 1$ and $2$ correspond to the magnetization and its second moment, respectively, and $\xi(t)$ denotes the
spatial correlation length. In simple cases, one has $\xi(t)\sim t^{1/z}$ with $z$ being the dynamic exponent,
and the magnetization is independent of $L$ in thermodynamic limit $L \to \infty$.
Then the scaling form of the vertical magnetization can be simplified as
\begin{equation}
M_\perp(t,x)=t^{-\eta /2z} \widetilde{M}_\perp(t^{1/z}/x).
\label{equ26}
\end{equation}
Inside the domain interface, i.e., $s=t^{1/z}/x \rightarrow \infty$, the scaling function obeys $\widetilde{M}_\perp(s) \sim s^{-\eta_0/2}$.
The vertical magnetization $M_\perp(t,x)$ then appears to exhibit a power-law behavior,
\begin{equation}
M_\perp(t,x) \sim t^{-(\eta+\eta_0)/2z} x^{\eta_0/2}.
\label{equ27}
\end{equation}
It decays much faster than that at bulk, i.e., $s \rightarrow 0$, where $\widetilde{M}_\perp(s) $ remains constant, and $M_\perp(t,x)$ behaves as
\begin{equation}
M_\perp(t,x) \sim t^{-\eta/2z}.
\label{equ28}
\end{equation}
The bulk and interface exponents, $\eta/2=0.117(2)$ and $\eta_0/2=0.997(7)$, have already been measured accurately in Ref.\cite{he09}.

Unexpectedly, the other component $M_\|(t,x)$ evolves quite differently. When the initial value of the $2\phi$
between the orientations of the two domains is strictly set to $\pi$, $M_\|(t,x)$ should always be zero according to the antisymmetry of the semi-ordered initial state. When $2\phi$ deviates only slightly from $\pi$, e.g. $0.988\pi$, the scaling form of $M_\|$ is expected
\begin{equation}
M_\|(t,x)=\xi(t)^{\psi z} \widetilde{M}_\|(\xi(t)/x)=t^{\psi}\widetilde{M}_\|(t^{1/z}/x),  \label{equ30}
\end{equation}
where the exponent $\psi$ is introduced to characterize the abnormal increase of the horizontal magnetization. At bulk, however, it decays with the time $M_{\|}(t,x) \sim t^{-\eta/2z}$, the same as that of the vertical one $M_{\perp}(t,x)$. On the right side of Eqs.~(\ref{equ26}) and (\ref{equ30}), both of the overall factors, $t^{-\eta /2z}$ and $t^{\psi}$, indicate the scaling dimensions of the two components of $\vec{M}(t, x)$ inside the domain interface, and the scaling function $\widetilde{M}_{\perp,\|}(\xi(t)/x)$ represents the scale invariance of the dynamic system. In general, they hold in the {\it macroscopic} short-time regime \cite{zhe03,zho08,zho09,he09,zho14}, after a microscopic time scale $t_{\rm mic}$ which is $100$ to $200$ Monte Carlo time steps (MCS) in this work.

Similarly, we may write down the dynamic
scaling form of the two-time correlation function,
\begin{equation}
C(t',t,x) = \xi(t')^{-\eta}\widetilde{C}\left (
\xi(t)/\xi(t'),\xi(t')/x \right ). \label{equ50}
\end{equation}
Since the scaling function depend on two scaling variables $\xi(t)/\xi(t')$ and $\xi(t')/x$, the dynamic behavior of
$C(t',t,x)$ is relatively complicated. Let us denote $s'=\xi(t')/x$ and $r=\xi(t)/\xi(t')$ for simplicity. Theoretically, in the large-$r$
limit, the scaling function $\widetilde{C}(s)$ is expected to exhibit power-law behavior both at bulk and inside the domain interface.
Careful analysis leads to the form
\begin{equation}
\widetilde{C}(r,s') \sim \left\{
   \begin{array}{lll}
    r^{-\lambda_b} ,     & \quad &  \mbox{ at bulk } \\
    r^{-\lambda_s},      & \quad &  \mbox{ inside domain interface}
   \end{array}, \right.
\label{equ40}
\end{equation}
where $\lambda_b=d+\eta/2$ and $\lambda_s=\eta_0/2-z\psi$ are the decay exponents for the bulk and domain interface, respectively.
In the small-$r$ regime, however, it shows a slight deviation from the power-law behavior. The logarithmic form of the correction should be considered in the growth of the spatial correlation length $\xi(t)$ due to the dynamic effect of the vortex-pair annihilation at KT transition
\cite{bra00,lei07},
\begin{equation}
\xi(t)\sim [t/(\ln t+c)]^{1/z}.
\label{equ60}
\end{equation}
In this paper, a more complicated correction form to scaling is taken,
\begin{equation}
\xi(t)\sim [t/(\ln t+c_1)]^{1/z}(1+c_2/t),
\label{equ65}
\end{equation}
where $c_1$ and $c_2$ are fitting parameters.

\section{Monte Carlo Simulation}

For the $2$D XY model, our main results are presented with $L = 512$
at $T = 0.89$, and the maximum updating time is $t_{\rm max} = 25~600$.
The total of samples for average is $10~000$.
The statistical errors are estimated by dividing the total samples
into two or three subgroups. If the fluctuation in the time
direction is comparable with or larger than the statistical error,
it will be taken into account.

\begin{figure*}[p]
\epsfysize=8cm \epsfclipoff \fboxsep=0pt
\setlength{\unitlength}{1.cm}
\begin{picture}(10,9)(0,0)
\put(-3.3, 0.0){{\epsffile{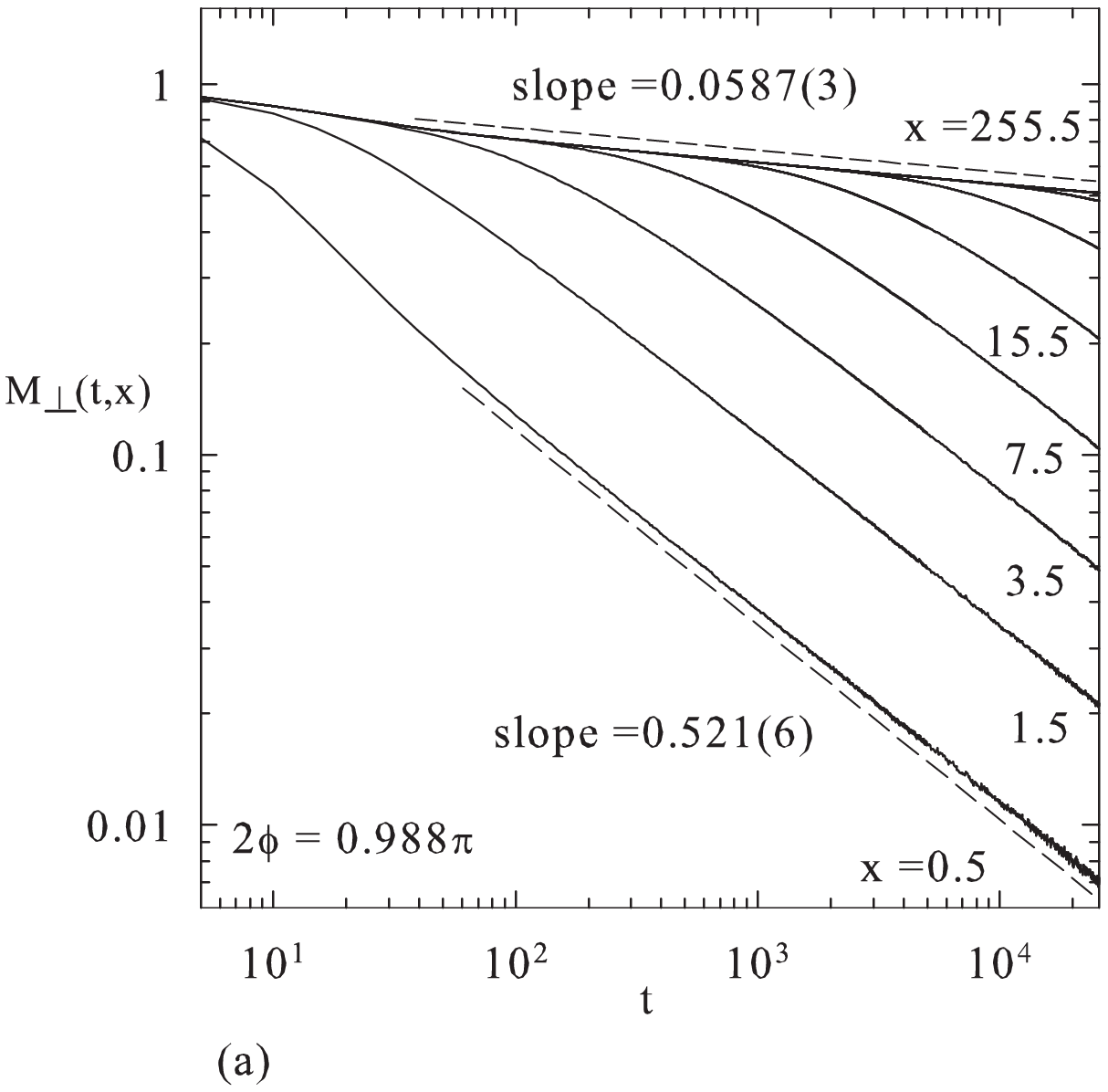}}}\epsfysize=8cm
\put(5.2, 0.0){{\epsffile{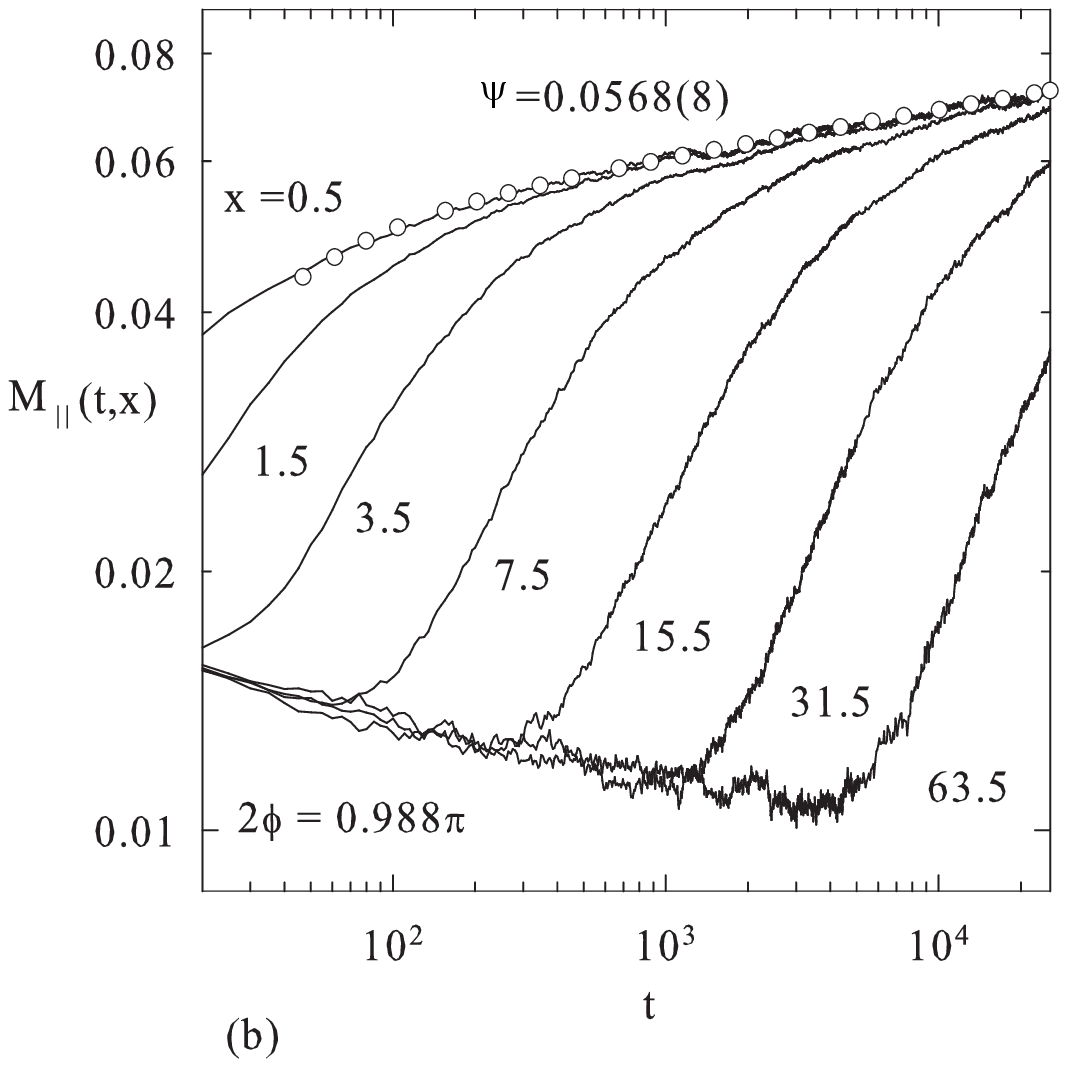}}}
\end{picture}
\caption{ Time evolution of magnetization ($M_\|, M_\perp$) starting from the semi-ordered state with the angle $2\phi = 0.988\pi$ on a double-log scale. Dashed lines show the power-law fits. The circles represent a power-law correction $M_\|\sim t^\psi(1+c/t)$ to scaling, with $\psi=0.0568$ as a result.
} \label{f2}
\end{figure*}

Firstly, we focus on the time evolution of the magnetization of $2$D
XY model starting from the semi-ordered states where the values of the angle $2\phi$ are very
close to $\pi$, including $0.984\pi, 0.986\pi, 0.988\pi, 0.990\pi$, and $0.992\pi$.
Only the results of $2\phi=0.988\pi$ are shown in this paper, and those of the
others behave quite similarly. As shown in Fig.~\ref{f2}(a), the vertical
component $M_\perp(t,x)$ seems to be the same as that starting from a
domain wall formed by the strictly opposite spins ($2\phi=\pi$) \cite{he09}.
According to Eqs.~(\ref{equ26})-(\ref{equ28}), $M_\perp(t,x)$ shows
the power-law decay for a sufficiently small $s$, e.g., $x=255.5$ and
$t<10000$, which makes $\widetilde{M}_\perp(s)$ a constant at bulk when $s\to 0$.
For a sufficiently large $s$, e.g., $x=0.5$ and $t>100$, $M_\perp(t,x)$
decays much faster, and we have $\widetilde{M}_\perp \sim s^{-\eta_0/2}$
inside the interface when $s\to \infty$. The critical exponent $\eta/2z=0.0587(3)$
measured from the slope of the curve at $x=255.5$ is well consistent with
$\eta=0.234(2)$ and $z=2$ reported in the literature \cite{tom02,zhe03}. And the other exponent
$(\eta + \eta_0)/2z=0.521(6)$ from the case at $x=0.5$ gives $\eta_0/2=0.93(1)$ by taking $z=2$ as input.
Considering there exists a strong correction to scaling in the growth of $\xi(t)$ as shown in Eq.~(\ref{equ65}), one refines the interface exponent $\eta_0/2=1.00(2)$, in good agreement with the results in Refs.\cite{zho08,he09}.

The horizontal component of the magnetization $M_\|(t,x)$ behaves quite
differently from $M_\perp(t,x)$. For the case $2\phi=0.988\pi$, $M_\|(t,x)$ shows
a tendency of the power-law increase at $x=0.5$ in Fig.~\ref{f2}(b).
Direct measurement of the curve after $t>200$ gives
the exponent $\psi=0.0572(6)$ in Eq.~(\ref{equ30}), and a fit to the numerical data with a power-law correction
extends to early times very well yielding $\psi=0.0568(8)$.
Besides, the influence of $2\phi$ on the exponent $\psi$ is also investigated.
Notice that the closer $2\phi$ is set to $\pi$, the better power-law increase of $M_\|$
there should be. However, the value too close to $\pi$ will result in strong fluctuations on $\vec{M}(t,x)$.
In this paper, $\psi=0.057(1)$ is confirmed by different values of $2\phi$ varying from  $0.986\pi$ to
$0.992\pi$, supporting the theoretical result $\psi=\eta/2z$ in Sec.~IV. Additionally, the curves of $x=127.5$ and $x=255.5$ simply obey the scaling form of $M_\|(t,x)$ at bulk, the same as that of $M_\perp(t,x)$ in Eq.~(\ref{equ28}). Thus they are omitted in
Fig.~\ref{f2}(b) due to the small value of $M_\|$ (e.g., $M_\|< 0.01$) comparable to the fluctuates.

\begin{figure*}[p]
\epsfysize=8.0cm \epsfclipoff \fboxsep=0pt
\setlength{\unitlength}{1.cm}
\begin{picture}(10,9)(0,0)
\put(-3.3,-0.0){{\epsffile{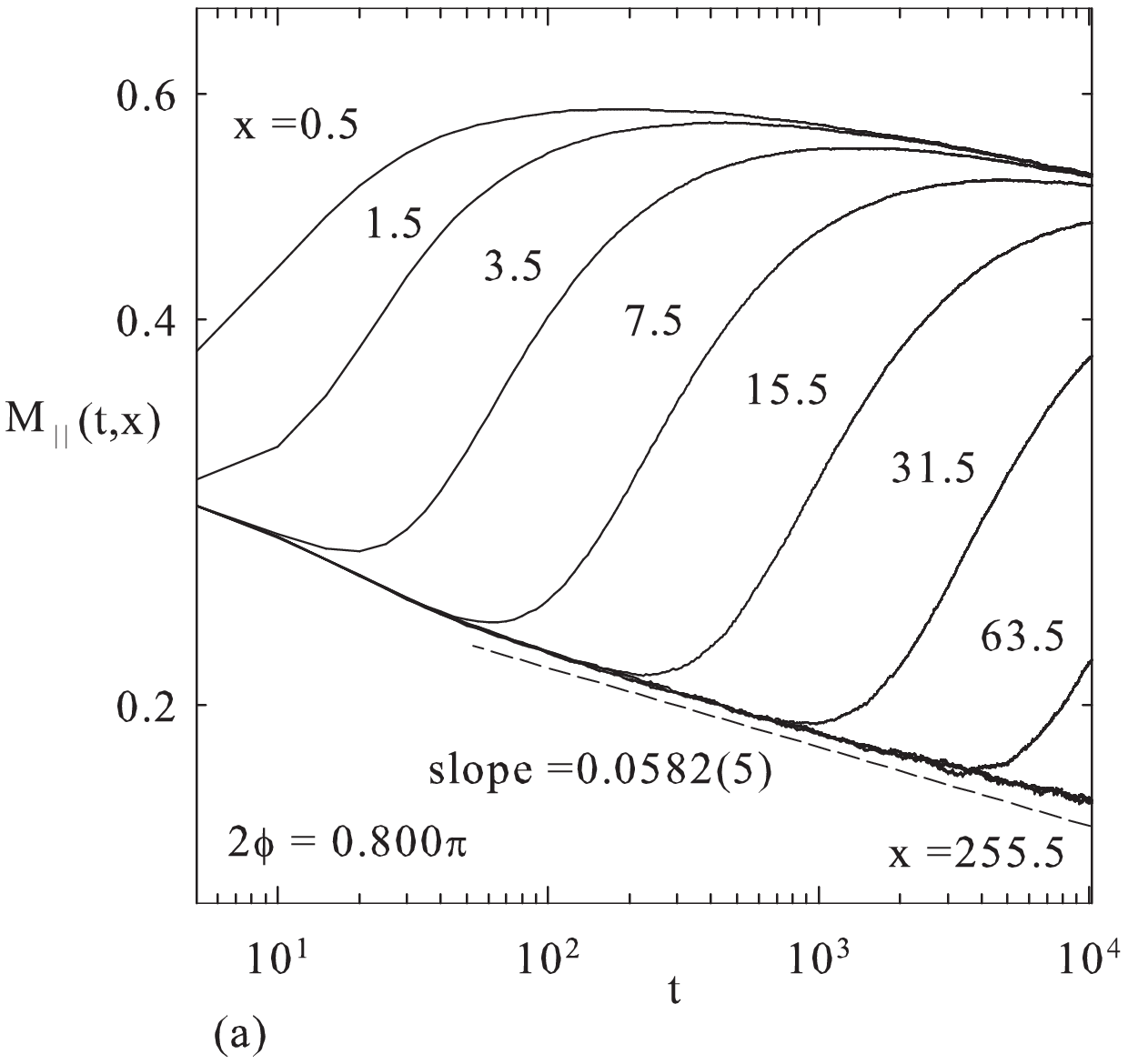}}}\epsfysize=8.0cm
\put(5.2,-0.0){{\epsffile{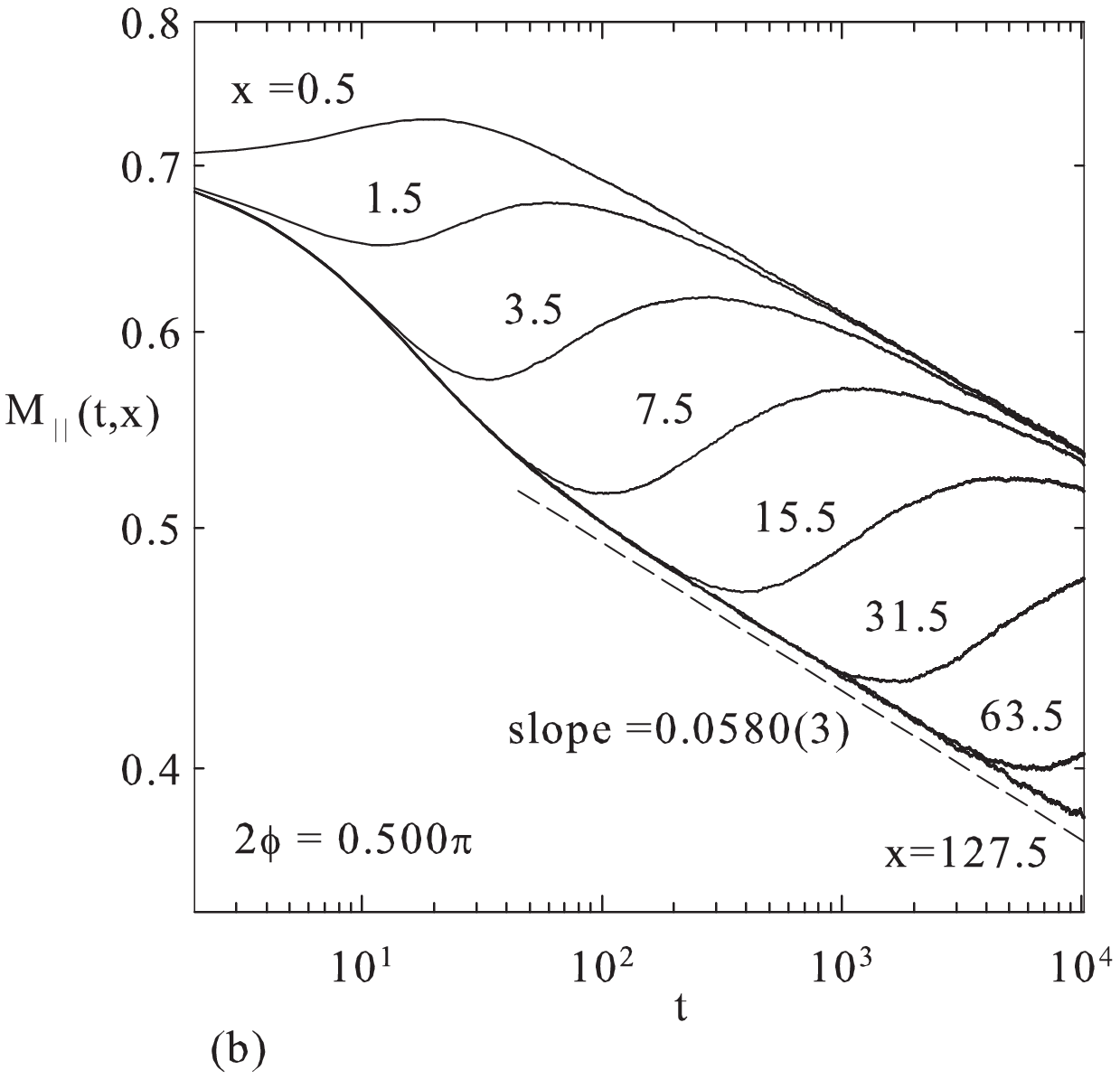}}}
\end{picture}
\caption{ The horizontal component of magnetization $M_\|$ as a function of the time $t$ for the initial states with the angle $2\phi=0.8\pi$ in (a) and $0.5\pi$ in (b). Dashed line show the power-law fits.}
\label{f3}
\end{figure*}

When the initial value of $2\phi$ is distinct from $\pi$, e.g., $2\phi = 0.8\pi$ and $0.5\pi$ in Fig.~\ref{f3}, the monotonic, pow-law increase of $M_\|(t,x)$ vanishes at $x=0.5$ inside the domain interface. A pow-law decay of $M_\|(t,x)$ occurs  after a crossover stage. The values of the exponent $\eta/2z =0.0582(5)$ and $0.0580(3)$ are measured from the slopes of the curves in the subfigures (a) and (b), respectively, again in agreement with that in the literature \cite{zhe03}.

\begin{figure*}[p]
\epsfysize=7.5cm \epsfclipoff \fboxsep=0pt
\setlength{\unitlength}{1.cm}
\begin{picture}(10,9)(0,0)
\put(-4.0,0.0){{\epsffile{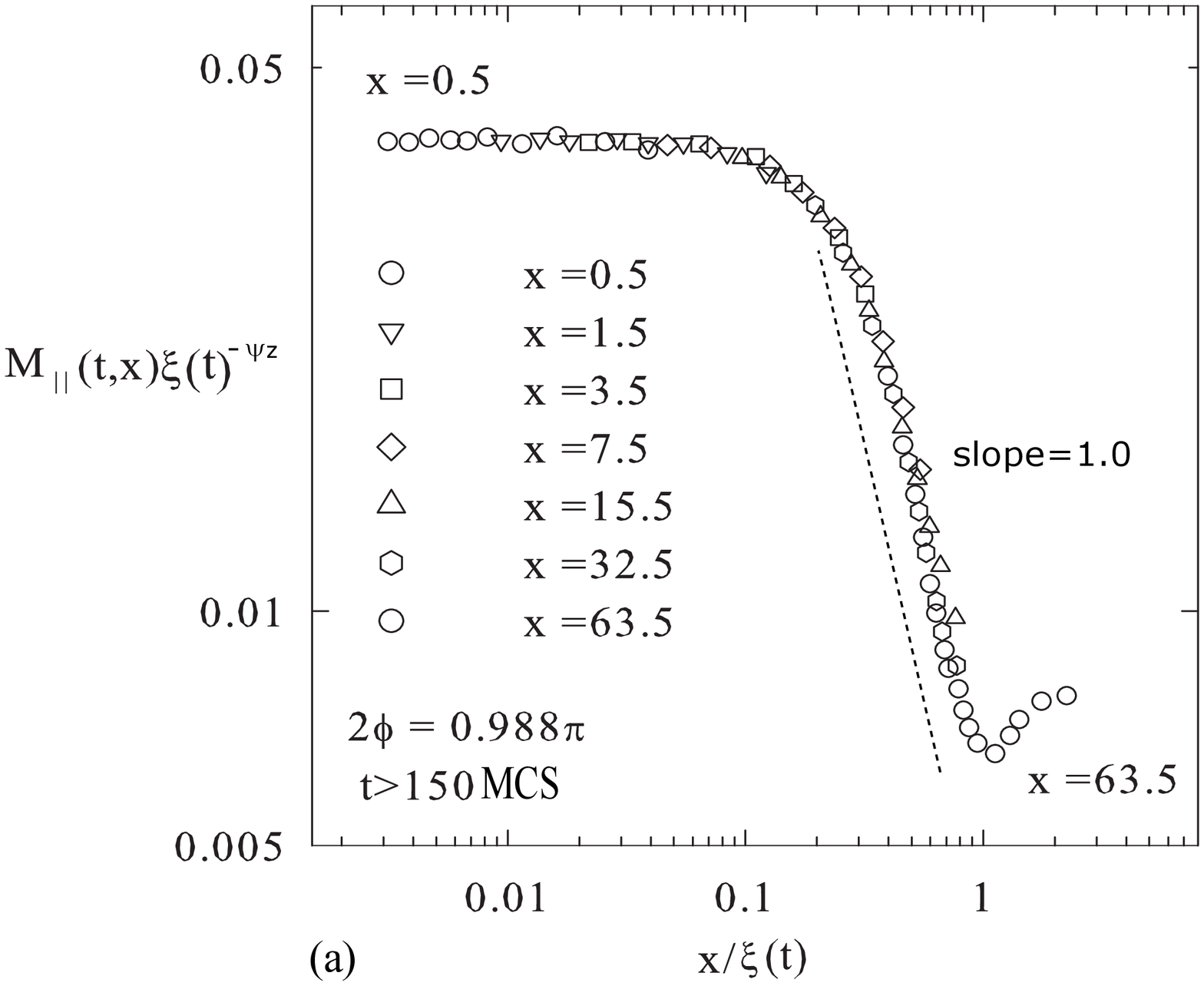}}}\epsfysize=7.7cm
\put(5.3,-0.1){{\epsffile{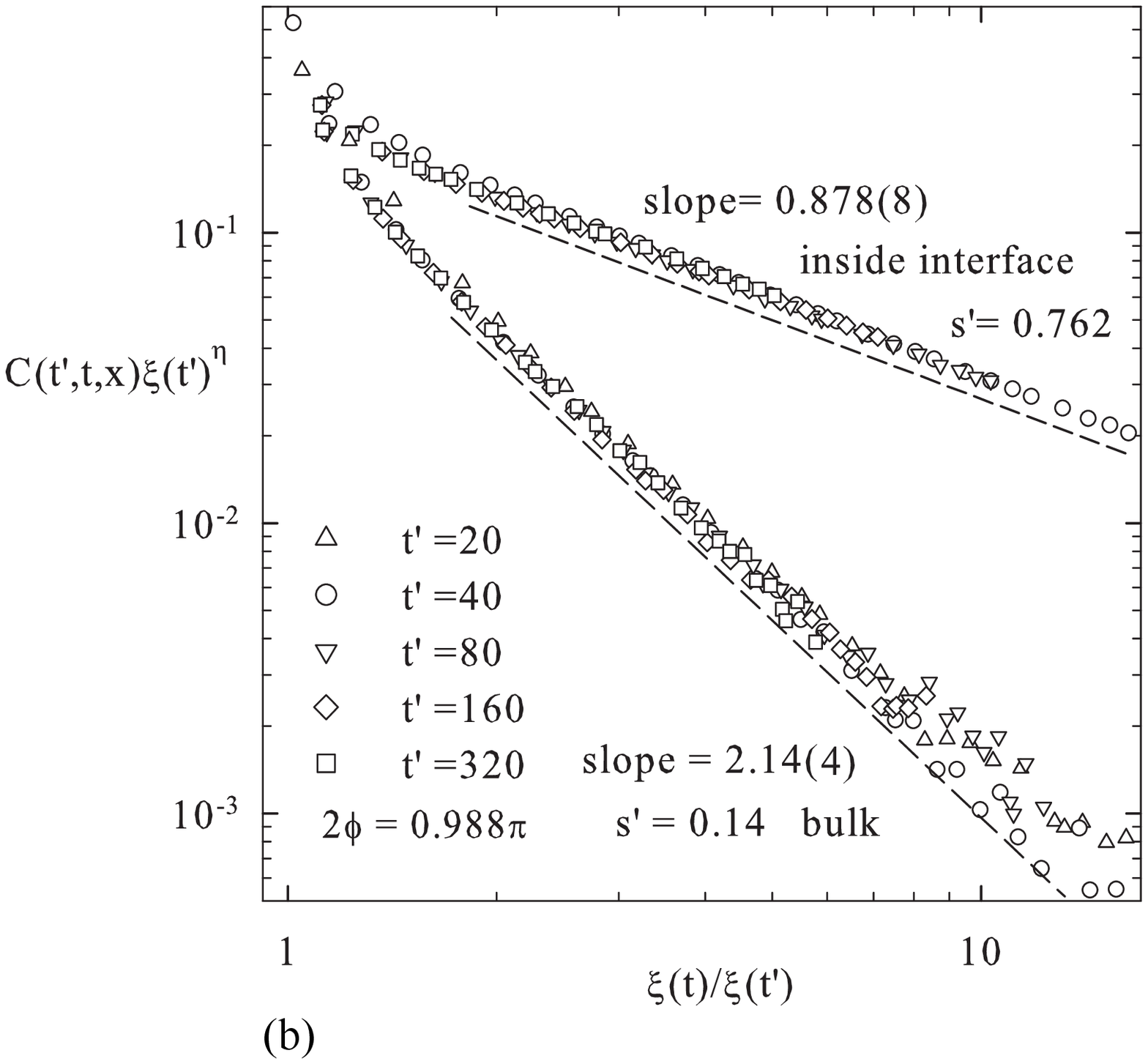}}}
\end{picture}
\caption{The scaling functions $M_\|(t,x)\xi(t)^{-\psi z}$ with respect to $x/\xi(t)$ in (a) and $C(t',t,x)\xi(t')^{\eta}$ with a
fixed $s'=\xi(t')/x$ as a function of $\xi(t)/\xi(t')$ in (b) on a double-log scale. The initial states with the angles $2\phi=0.998\pi$
is prepared, and data collapse for different $x$ and $t'$ are observed when the correction to the scaling defined in Eq.~(\ref{equ65}) is considered for the correlation length $\xi(t)$. Dashed lines show the power-law fits.
}
\label{f4}
\end{figure*}

In Fig.~\ref{f4}(a), the scaling function of $\widetilde{M}_\|(t,x)$ defined in
Eq.~(\ref{equ30}) is plotted as a function of $x/\xi(t)$ for the horizontal component
of magnetization $M_\|(t,x)$ with the initial value of the angle $2\phi=0.988\pi$,
where the correlation length $\xi(t)$ is calculated according to Eq.~(\ref{equ65}) with $c_1=5.45$ and $c_2=-9.1$.
Data of different $x$ collapse clearly onto the master curve at $t>150$ MCS.
Inside the domain interface, $\widetilde{M}_\|(x/\xi(t)) \rightarrow  \rm {const}$
is observed in the limit $x/\xi(t) \rightarrow 0$, different from that of
the vertical component $M_\perp(t,x)$ where $\widetilde{M}_\perp(s) \sim s^{-\eta_0/2}=(x/\xi(t))^{\eta_0/2}$.
An increase of $\widetilde{M}_\|(x/\xi(t))$ is then observed at $x/\xi(t)>1$ with the slope $0.23(1)$, leading to $\eta\approx 0.23$
from the usual expectation $M_\|(t) \sim \xi(t)^{-\eta/2}$ at bulk, comparable with the earlier results \cite{zhe03}. Between them, it exhibits a power-law decrease with the slope $\eta_0/2=1.00(7)$.

The behavior of two-time correlation function is also carefully examined.
After subtracting the contribution of the
magnetization, $C(t',t,x)$ describes the pure time
correlation. As shown in Fig.~\ref{f4}(b), the scaling variable $s'=\xi(t')/x$ is
fixed at certain values, e.g., $s'=0.14$ and $0.762$ for the bulk and domain interface, respectively.
In both of the two cases, the function $C(t',t,x)\xi(t')^\eta$ shows data collapse with respect to $\xi(t)/\xi(t')$. Since $C(t',t,x)$ at bulk decays rapidly, the data is relatively fluctuating. The slopes of the power-law tails give the exponents
$\lambda_b=d+\eta/2=2.14(4)$ and $\lambda_s=\eta_0/2-\psi z =0.878(8)$ in Eq.~(\ref{equ40}).
Thus one has $\eta/2=0.14$ and $\psi=0.061$, comparable with those obtained from the magnetization in Fig.~\ref{f3}.

\section{Theoretical Analysis}

In this section, the dynamic behavior of the horizontal and vertical components of the magnetization are
analyzed in the long-wavelength approximation which is a conventional treatment on the KT phase transition 
of the two-dimensional XY model. Recently this approximation is also used to investigate the non-equilibrium 
critical dynamics \cite{ber01,jel11}.  In general, it is valid in the low-temperature regime, well below the 
critical temperature $T_{c}$. Near $T_c$ there may exist the dynamic effect of the vortices. In our theoretical analysis, 
however, the initial states are ordered and semi-ordered states, and in both dynamic processes, the vortex effect is suppressed. The dynamical behaviors seem similar to those of the spin waves with very long wavelengths. 
\subsection{Ordered initial state}
With the long-wavelength approximation, the Hamiltonian of the XY model then can be rewritten as \cite{kos72,kap07,ber01}
\begin{eqnarray}
\mathcal{H} &=& - \sum_{<ij>}\cos(\theta_i - \theta_j)\nonumber \\
& \approx & \frac{1}{2}\sum_{<ij>}\left(\theta_i - \theta_j \right)^2  + \mathcal{H}_0 \nonumber \\
& = &  \frac{1}{4}\sum_{\vec R} \sum_{\vec a} \left(\theta(\vec R) -
\theta(\vec R + \vec a) \right)^2 + \mathcal{H}_0.
\label{equ120}
\end{eqnarray}
Where $\theta_i$ is the orientation angle of the spin at the site $i$,
$\vec R$ is the position vector in the plane, $\vec a$ is the unit vector between the site and its nearest neighbors, and
the constant term $\mathcal{H}_0 \approx -1$ does not affect the dynamics at all. After the
fourier transformation of $\theta(\vec R)$, one obtains an effective Hamiltonian,
\begin{equation}
\mathcal{H}_{\rm eff} = \frac{1}{2}\sum_{\vec k}J(\vec k)\left|\theta(\vec k)\right|^2,
\label{equ130}
\end{equation}
where $|\cdots| $ represents the modulus, and the function $J(\vec k)$ satisfies
\begin{equation}
J(\vec k) = \frac{1}{2}\sum_{\vec a} \left| 1 - e^{i\vec k \cdot
\vec a }\right|^2 \approx k^2a^2. \label{equ140}
\end{equation}
Note that the wave-vector $\vec{k}$ is two-dimensional, ranging from $-\pi/a$ to $\pi/a$ in any direction.
For convenience, the notation of the vector is omitted in the following.

The dynamics of the XY model is investigated with the Langevin equation in the momentum space \cite{kim97,ber01},
\begin{eqnarray}
\frac{d \theta(k, t)}{dt} & = & - \frac{\rho_s}{T} \frac{\partial \mathcal{H}_{\rm eff}}{\partial \theta(k, t)} + \epsilon(k,t) \nonumber \\
& = & -\frac{a^2k^2\rho_s}{T}\theta(k, t) + \epsilon(k, t). \label{equ150}
\end{eqnarray}
Where $T$ is the temperature of the system, $\epsilon(k, t)$ is Gaussian white
noise with the correlation given by the fluctuation-dissipation theorem
$\langle \epsilon(k, t) \epsilon(k', t')\rangle = 2 \delta(k + k') \delta(t
- t')$, and the spin-wave stiffness $\rho_s = 1$ is set. The above linear equation can be solved as
\begin{equation}
\theta(k, t) = \int_{0}^{t}e^{-a^{2}k^{2}(t-t')/T}\epsilon(k,
t') dt' + \theta(k, 0) e^{-a^{2}k^{2}t / T }.
\label{equ160}
\end{equation}
The vanishing value of $\langle\theta(k, t)\rangle = \theta(k, 0) e^{-a^{2}k^{2}t/ T}$ is expected in equilibrium with
$t \rightarrow \infty$. Furthermore, its second moment are also calculated with Eq.~(\ref{equ160}),
\begin{equation}
\langle \left|\theta(k, t)\right|^{2}\rangle =
\frac{T}{a^2k^2}\left(1-e^{-2a^{2}k^{2}t/T}\right) +
\left|\theta(k, 0)\right|^{2}e^{-2a^{2}k^{2}t/T}.
\label{equ170}
\end{equation}
The result of the equilibrium state $\langle \left|\theta(k, t)\right|^{2}\rangle = T / (a^2k^2)$
is then obtained, the same as that obtained from the equipartition theorem in equilibrium
based on Eqs.~(\ref{equ130}) and (\ref{equ140}).

Formally, the solution $\theta(k, t)$ in Eq.~(\ref{equ160})can be divided into two parts, the bulk one $F(k, t)$ and the initial one
$G(k, t)$. The magnetization $M(t, R)$ is then calculated with $F(R, t)$ and $G(R, t)$ obtained by the inverse fourier transformation,
\begin{equation}
M(t, R) = \left \langle e^{i\theta(R, t)} \right \rangle  = \left
\langle e^{iF(R,t)} \right \rangle e^{iG(R, t)}.
\label{equ190}
\end{equation}
Using the accumulator variable expansion and the vanishing expected value $\langle F(R,t)\rangle=0$, the time evolution of the magnetization $\vec{M}=(M_\|, M_\perp)$ is deduced as
\begin{eqnarray}
M_\|(t, R) &\approx& e^{- \langle F^2(R, t)\rangle /2 } \cos(G(R, t)), \nonumber \\
M_\perp(t, R) &\approx & e^{- \langle F^2(R, t)\rangle /2 } \sin(G(R, t)).
\label{equ200}
\end{eqnarray}
Since the equal-time correlation function at bulk satisfies
\begin{equation}
\langle F(k,t)F(k',t)\rangle  = \frac{\delta(k + k') T}{a^{2}k^{2}}\left(1-e^{-2a^{2}k^{2}t/T} \right),
\label{equ210}
\end{equation}
the function $\langle F^2(R,t)\rangle$ in Eq.~(\ref{equ200}) can be calculated by the inverse fourier transformation,
\begin{eqnarray}
& & \langle F^2(R,t)\rangle \nonumber \\
&& = \left(\frac{a}{2\pi}\right)^{4}\int_{-\infty}^{\infty} dk\int_{-\infty}^{\infty} dk'\left\langle F(k,t)F(k',t)\right\rangle e^{i\left(k+k'\right)\cdot R} \nonumber \\
&& =\left(\frac{a}{2\pi}\right)^{2}\int_{-\infty}^{\infty} dk\frac{T}{a^{2}k^{2}}\left(1-e^{-2a^{2}k^{2}t/T}\right),
\label{equ220}
\end{eqnarray}
where $a / 2\pi$ is a normalization factor, and $k$ is the two-dimensional vector in the momentum space. The above
integration is calculated as
\begin{eqnarray}
& & \langle F^2(R,t)\rangle \nonumber \\
& & =\left(\frac{a}{2\pi}\right)^{2}\iint_{-\infty}^{\infty} dk_x dk_y \frac{T}{a^{2}(k_x^2+k_y^2)}\left(1-e^{-2a^{2}(k_x^2+k_y^2)t/T}\right) \nonumber \\
& & = \left(\frac{a}{2\pi}\right)^{2} \left(\frac{T\pi}{a^2}\ln t + \rm {C}_1\right) \nonumber \\
& & = \frac{T}{4\pi}\ln t + \rm{C}_2,
\label{equ225}
\end{eqnarray}
where $\rm C_1$ and $\rm C_2$ are integral constants. Note that the function $\langle F^2(R, t)\rangle$ is independent of the position vector $R$,
suggesting that bulk part of the orientation angle $\theta(R, t)$ is uniform in the plane.

On the other hand, the initial part $G(R,t)$ is obtained with
\begin{eqnarray}
G(R, t)& = & \left(\frac{a}{2\pi}\right)^2\int_{-\infty}^{\infty} dk e^{ik\cdot R}G(k,t) \nonumber \\
       & = & \left(\frac{a}{2\pi}\right)^2\int_{-\infty}^{\infty} dk e^{ik\cdot R} \theta(k,0) e^{-a^{2}k^{2}t/T},
\label{equ230}
\end{eqnarray}
where $\theta(k,0)$ is the Fourier transform of the initial value of the spin orientation $\theta(R, 0)$,
\begin{equation}
\theta(k, 0) = \int_{-\infty}^{\infty} dR' e^{-ik\cdot R'} \theta(R', 0).
\label{equ235}
\end{equation}
In the case with the ordered initial state with $\theta(R, 0) = \phi$,
$\theta(k, 0)=4\pi^2\phi\delta(k)/a^2$ and $G(R, t) = \phi$ are calculated with
Eqs.~(\ref{equ235}) and (\ref{equ230}), respectively. Substituting the functions $\langle F^2(R, t)\rangle$ and
$G(R,t)$ into Eq.~(\ref{equ200}), the time evolution of the magnetization is then derived analytically,
\begin{eqnarray}
M_\|(t, R) & \propto & t^{- T/8\pi}\cos(\phi), \nonumber \\
M_\perp(t, R) & \propto & t^{- T/8\pi}\sin(\phi).
\label{equ240}
\end{eqnarray}
Note that the above analysis is based on the long-wavelength approximation which is valid only in the low-temperature phase.
Therefore, the power-law decay of the magnetization holds at $T\leq T_c$. In Refs. \cite{bra00, kap07, jel11},
$\eta(T) = 1/(2 \pi \beta J) = T/(2\pi)$ and $z = 2$ were reported in the $2$D XY model. The dynamic behavior of the magnetization $M \propto t^{- \eta / 2 z}$ is then deduced with the ordered initial state, the same as that from the short-time dynamic scaling theory \cite{zhe03}.

\subsection{Semi-ordered initial state}
The above analysis on the critical dynamics starting from the ordered state has been confirmed to be valid though it is very crude.
What about the critical dynamics of the system with a semi-ordered initial state?
For simplification, we set the initial values of the spin orientations $\theta(\vec R, 0) =\phi \epsilon(x)$, where $x$ is one of the space components defined in the direction
perpendicular to the perfect domain wall, $2\phi$ is the angle within the interval $[0, \pi]$, and the
function $\varepsilon(x)$ is defined as
\begin{equation}
   \varepsilon(x) =  \left\{
   \begin{array}{lll}
   -1,   & \quad &  \mbox{if $x \leq 0$ } \\
    1,   & \quad &  \mbox{if $x > 0$ }
   \end{array}\right. .
   \label{equ450}
\end{equation}
Following Eqs.~(\ref{equ230}) and (\ref{equ235}), one can calculate the functions
$\theta(k,0)$, $G(k,t)$, and $G(R,t)$
\begin{eqnarray}
\theta(k, 0) & = & \int_{-\infty}^{\infty} dR e^{-i k \cdot R}\theta(R, 0) \nonumber \\
 & = & \phi\int_{-\infty}^{\infty} dx e^{-ik_x x }\varepsilon(x)\delta(k_y) \nonumber \\
 & = & \frac{4\pi\phi\delta(k_y)}{iak_x},
\label{equ460}
\end{eqnarray}
\begin{equation}
G(k,t) = \theta(k, 0)e^{-a^{2}k^{2}t/T} =
\frac{4\pi\phi\delta(k_y)}{iak_x}e^{-a^{2}k^{2}t/T},
\label{equ470}
\end{equation}
\begin{eqnarray}
G(R, t) & = & \left(\frac{a}{2\pi}\right)^2\int_{-\infty}^{\infty} dk e^{i k \cdot R}G(k,t) \nonumber \\
 & = & \left(\frac{a}{2\pi}\right)\int_{-\infty}^{\infty} dk_x e^{ik_x x}\frac{2\phi}{ik_x}e^{-a^{2}k_x^{2}t/T}.
\label{equ480}
\end{eqnarray}
By means of the derivative of the $G(R, t)$,
\begin{eqnarray}
\frac{\partial G(R, t)}{\partial x} & = & \left(\frac{\phi a}{\pi}\right)\int_{-\infty}^{\infty} dk_x e^{ik_x x}e^{-a^{2}k_{x}^{2}t/T}  \nonumber \\
 & = & \phi\sqrt{\frac{T}{\pi
 t}}e^{-(T/4t)\left(x/a\right)^{2}},
\label{equ490}
\end{eqnarray}
the initial function $G(x, t)$ can be solved as
\begin{equation}
G(x',t) = \phi\int_{0}^{x'} dr \sqrt{\frac{T}{\pi t}}e^{-(T/4t)r^{2}} + C_3.
\label{equ500}
\end{equation}
Here the integral constant $C_3 = 0$ is derived from the symmetry analysis, and $x' = x / a$ is a dimensionless number.
For convenience, we use $x$ instead of $x'$ to denote the value of the position in the $x$ direction.
The initial function $G(x,t)$ is further simplified with the notation $s = (\sqrt{T}/2) x /t^{1/z}$ being the ratio between the
position $x$ and the correlation length $\xi(t) \sim t^{1/z}$ wherein $z=2$,
\begin{equation}
G(s)=\frac{2\phi}{\sqrt{\pi}}\int_{0}^{s}ds' e^{-s'^{2}}.
\label{equ510}
\end{equation}
The conclusion that $G(s)$ is a function of the single variable $s$ is quite
consistent with that obtained from the scaling arguments in
Refs.~\cite{zho08,he09}. Finally, the dynamic behavior of the magnetization
$M = (M_\|, M_\perp)$ is identified,
\begin{eqnarray}
M_{\|}(t, x) & \propto & t^{-\eta / 2z} \cos \left(\frac{2\phi}{\sqrt{\pi}}\int_{0}^{s}ds'e^{-s'^{2}}\right), \nonumber \\
M_{\perp}(t, x) & \propto & t^{-\eta / 2z} \sin
\left(\frac{2\phi}{\sqrt{\pi}} \int_{0}^{s}ds'e^{-s'^{2}}\right),
\label{equ520}
\end{eqnarray}
with $s= (\sqrt{T}/2) x /t^{1/z}=\sqrt{\pi\eta/2}~x /t^{1/z}$ at the KT transition.
There are two different regimes of $\vec M(t, x)$. At bulk the magnetization exhibits a power-law decay $t^{-\eta / 2z}$,
just the same as that of the ordered initial state. Inside the domain interface, however, the behavior of
the magnetization is quite different. Using $G(s) = 2\phi s/\sqrt{\pi}$, one obtains
\begin{eqnarray}
M_{\|}(t, x) & \propto & t^{-\eta / 2z }\left(1-\frac{1}{2}G^2(s)\right) \nonumber \\
 & \propto & t^{-\eta / 2z }\left(1-\frac{1}{2}\frac{4\phi^{2}s^2}{\pi}\right) \nonumber\\
 & \propto & t^{-\eta / 2z }\left(1- \frac{\phi^{2}x^2\eta}{t^{2/z}}\right),
\label{equ530}
\end{eqnarray}
\begin{eqnarray}
M_{\perp}(t, x) & \propto & t^{-\eta / 2z}G(s) \nonumber \\
 & \propto & \phi \sqrt{\frac{T}{\pi}} x t^{-1 / z - \eta/ 2 z}.
\label{equ540}
\end{eqnarray}

\begin{figure*}[p]
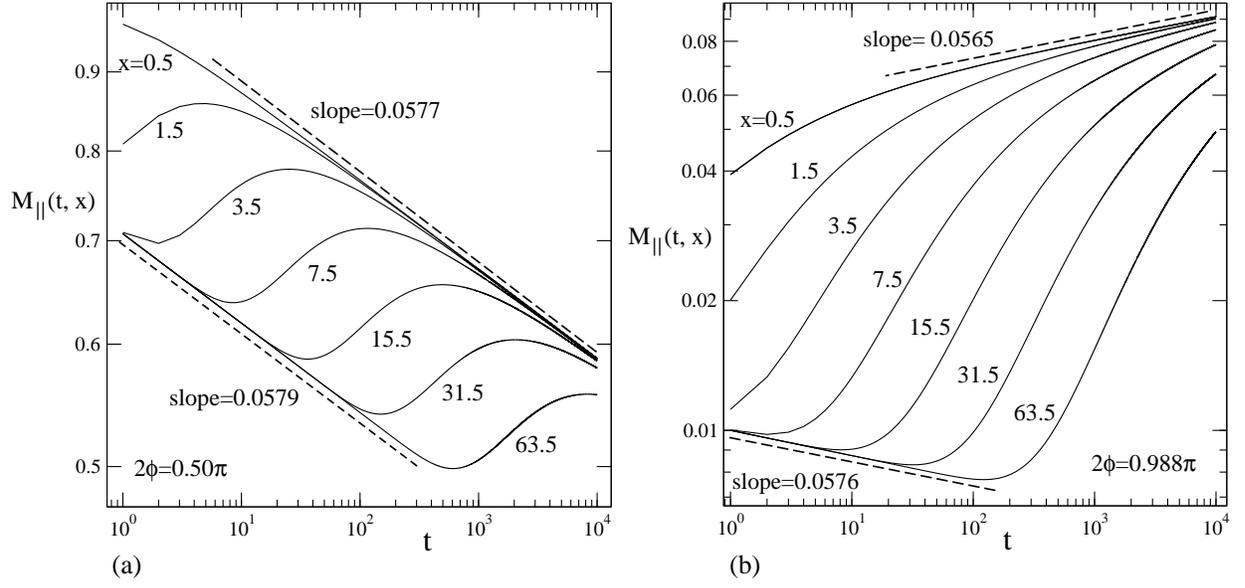

\epsfysize=7.7cm \epsfclipoff \fboxsep=0pt
\setlength{\unitlength}{1.cm}
\begin{picture}(10,9)(0,0)
\put(-3.0,-0.0){{\epsffile{fig05a.eps}}}\epsfysize=7.7cm
\put(5.2,-0.0){{\epsffile{fig05b.eps}}}
\end{picture}
\caption{ Theoretical results of the horizontal magnetization $M_\|(t, x)$ from analytical calculations based on the Langevin dynamics, as shown in   Eqs.~(\ref{equ520}) and (\ref{equ640}), are plotted at the angles $2\phi=0.50\pi$ in (a) and $0.988\pi$ in (b), respectively, for various values of $x$ as a function of the time $t$ on a log-log scale. The dashed lines represent the power-law fits.
}
\label{f5}
\end{figure*}

The behavior of $M_{\perp}(t, x)$ in Eq.~(\ref{equ540}) is in good agreement with simulation results in Ref.~\cite{he09} and
in Fig.~\ref{f2}(a) of this work. While the case of the other magnetization component, $M_{\|}(t, x)$, is quite complicated. When the initial value of the angle $2\phi$ is far less than $\pi$, for example $2\phi = 0.50\pi$, the critical behaviors of $M_{\|}(t, x)$ from Eq.~(\ref{equ520}) as displayed in Fig.~\ref{f5}(a), is quite consistent with those in Fig.~\ref{f3}(b) obtained from Monte Carlo simulations.
Especially, the slopes $0.0577$ and $0.0579$ are measured from the upper and lower envelopes, almost the same as $\eta/2z=0.0580(3)$ within errors.
When $2\phi$ is closer to $\pi$, such as $2\phi = 0.988\pi$, an abnormal increase of $M_{\|}(t, x)$ occurs at $x=0.5$ as shown in Fig.~\ref{f2}(b). Exactly at $2\phi = \pi$, $M_{\|}(t, x)\equiv 0$ is obtained from numerical simulations. Both of them show a visible deviation from the theoretical prediction in Eq.~(\ref{equ520}), pointing out the failure of the long-wavelength approximation.

\subsection{ Correction to long-wavelength approximation}
Since the long-wavelength assumption in Eq.~(\ref{equ120}) is invalid when $2\phi \approx \pi$,
the correction should be considered for the spins inside the domain interface,
\begin{equation}
\cos(\theta_i - \theta_j - \pi) \approx 1 - \frac{1}{2}(\theta_i - \theta_j - \pi)^2.
\label{equ550}
\end{equation}
The corresponding Hamiltonian $\mathcal{H}_s$ is then rewritten as
\begin{eqnarray}
\mathcal{H}_s &=& - \sum_{<ij>} \cos(\theta_i - \theta_j) \nonumber \\
  &=&  \sum_{<ij>} \cos(\theta_i - \theta_j - \pi) \nonumber \\
  & \approx & \sum_{<ij>} \left( 1 - \frac{1}{2}\left(\theta_i - \theta_j - \pi \right)^2 \right) \nonumber \\
& = & H_0 - \frac{1}{4}\sum_{R} \sum_{a} \left(\theta(R) -
\theta(R + a) - \pi \right)^2.
\label{equ560}
\end{eqnarray}
After a linear transformation
\begin{eqnarray}
\theta'( R ) &=& \theta( R ) - \frac{\pi}{2}, \nonumber \\
\theta'( R + \vec a) &=& \theta(R + a) + \frac{\pi}{2},
\label{equ570}
\end{eqnarray}
one renews the Hamiltonian of the domain interface
\begin{equation}
\mathcal{H}'_{s}= - \frac{1}{4}\sum_{R} \sum_{a} \left(\theta'(R) -
\theta'(R + a) \right)^2 + \mathcal{H}'_{0}.
\label{equ580}
\end{equation}
Comparing with Eq.~(\ref{equ120}), nothing is changed except for the sign of the first term on the right-hand side.
With the Langevin equation in Eq.~(\ref{equ150}) and the revised Hamiltonian, the dynamics of the domain interface
is carefully investigated. Similar with Eqs.~(\ref{equ225}) and (\ref{equ510}), one can deduce
\begin{eqnarray}
\langle F'^2(R,t)\rangle & = & - \frac{T}{4\pi}\ln t + C_4,   \nonumber  \\
G'(s)& =& \frac{2i\phi}{\sqrt{\pi}}\int_{0}^{s} ds' e^{s'^{2}},
\label{equ610}
\end{eqnarray}
where $C_4$ is the integral constant, and the variable $s = \sqrt{\pi\eta/2}~x /t^{1/z}$.
Hence, the correction to the long-wavelength approximation yields the dynamic behavior of magnetization in the limit of $s \rightarrow 0$,
\begin{eqnarray}
M_\|(t, x) & \propto & t^{ T/8\pi}\exp\left(i\frac{2i\phi}{\sqrt{\pi}}s\right)  \nonumber \\
 & \propto & t^{\eta/2z }\exp \left(- \phi\sqrt{2\eta} \frac{x}{t^{1/z}}\right).
\label{equ630}
\end{eqnarray}
In the above analysis inside the domain interface, one neglects the contribution from the bulk decay $M_\|(t, x) \propto t^{-\eta/2z}$, since it is
relatively small compared to the abnormal increase. On the contrary, outside the domain interface, i.e., in the limit of $s \rightarrow \infty$,
the decay behavior of the horizontal magnetization at bulk becomes dominant. Therefore one may write
\begin{equation}
M_\|(t, x) = A_1 t^{\eta/2z }\exp \left(- \phi\sqrt{2\eta} \frac{x}{t^{1/z}}\right) + A_2 t^{-\eta/2z},
\label{equ640}
\end{equation}
where $A_1$ and $A_2$ are coefficients of the linear superposition. Taking the initial value $2\phi=0.988\pi$ as an example, the critical behavior of the horizontal magnetization as described in  Eq.~(\ref{equ640}) is displayed in Fig.~\ref{f5}(b) for different values of the $x$, with the parameters $A_1=0.05, A_2=0.01, \eta/2z=0.0587$ and $\eta=0.234$ as input. The theoretical results agree characteristically with Monte Carlo results in Fig.~\ref{f2}(b). In particular, the slope $0.0565$ is measured from the increase of $M_\|(t, x=0.5)$, consistent with $\psi=0.0568(8)$, further supporting the relation $\psi=\eta/2z$. At bulk, $M_\|(t, x)$ approaches to the nonlinear decay with the slope $0.0576$, in well agreement with the expectation.

\section{Conclusion}
With Monte Carlo simulations and theoretical analyses based on the Langevin equation, the spin-reorientation critical dynamics starting from the semi-ordered initial states has been investigated, taking the $2$D XY model as an example. At the KT phase transition, dynamic scaling behaviors
of the magnetization containing two orthogonal components $M_{\perp}(t,x)$ and $M_{\|}(t,x)$ as well as the two-time correlation function $C(t',t,x)$ are carefully analyzed, and critical exponents are accurately determined. When the initial value of the angle $2\phi$ between the two directions of the adjacent domains is slightly lower than $\pi$, an abnormal power-law increase of $M_{\|}(t,x)$ is observed inside the domain interface, other than the well-known decay phenomenon at bulk.
The corresponding critical exponent $\psi=0.0568(8)$ is measured. Furthermore, the relation $\psi=\eta/2z$ is analytically deduced from the Langevin dynamics in the long-wavelength approximation, well consistent with the numerical results. When the initial value of $2\phi$ is much smaller than $\pi$, such as $0.5\pi$ and $0.8\pi$, however, $M_{\|}(t,x)$ obeys a power-law decay, instead of the increase.

Interestingly, a similar increasing behavior of the magnetization $M(t,m_0)\sim m_0t^\theta$ has been reported for the $2$D XY model starting from a disordered state with a small initial value $m_0$ \cite{zhe03}. Monte Carlo simulations at the KT phase transition gives an independent critical exponent $\theta = 0.241$, much larger than the value of $\psi$ in our work. Future studies are needed to identify the relation between these two exponents $\theta$ and $\psi$.

{\bf Acknowledgements:} This work was supported in part by National Natural Science Foundation of China (under
Grant Nos. $11775186$ and $11875120$) and Zhejiang Provincial Natural Science Foundation of China (under Grant No. LY$17$A$050002$).

\bibliography{xy_ref}
\bibliographystyle{apsrev4-1}
\newpage

\end{document}